\documentclass[12pt,draftcls,onecolumn]{IEEEtran}

\ifCLASSINFOpdf
\else
\fi

\ifCLASSOPTIONcompsoc
\usepackage[nocompress]{cite}
\else
\usepackage{cite}
\fi

\usepackage{amsmath}
\usepackage{amssymb}
\usepackage{graphicx}
\usepackage{epstopdf}
\usepackage{mathtools}
\usepackage{xcolor}

\begin{document}

\title{$\epsilon$-Arithmetics for Real Vectors and
  Linear Processing of Real Vector-Valued Signals with Real Vector-Valued
Coefficients}


\author{
        Xiang-Gen Xia, \IEEEmembership{Fellow}, \IEEEmembership{IEEE} 

\thanks{
X.-G. Xia is with the Department of Electrical and Computer Engineering, University of Delaware, Newark, DE 19716, USA (e-mail: xxia@ece.udel.edu).}

}

\date{}

\maketitle


\begin{abstract}
In this paper, we introduce a new concept, namely $\epsilon$-arithmetics, 
for real vectors of any fixed dimension. The basic idea is to use vectors of 
rational values (called rational vectors) to approximate vectors of real values of the same dimension within 
$\epsilon$ range. For rational vectors of a fixed  dimension $m$, they can form a 
field that is  an $m$th order extension $\mathbf{Q}(\alpha)$
of the rational field $\mathbf{Q}$ where $\alpha$ has its minimal 
 polynomial of 
degree $m$ over $\mathbf{Q}$. Then, the arithmetics, such as addition, 
subtraction, multiplication, and division, of real vectors can be defined by using that of their approximated rational vectors within $\epsilon$ range. 
We also  define complex conjugate of a real vector and then inner
product and convolutions of two real vectors and two real vector sequences
(signals) of finite length. With these newly defined concepts for real vectors,
linear processing, such as linear filtering, ARMA modeling, and  least squares 
fitting, with real vector-valued coefficients can be implemented
to real vector-valued signals, which will broaden the existing linear
processing to scalar-valued signals. 
\end{abstract}

\begin{IEEEkeywords}
\textit{Arithmetics, rational vectors, real vectors, rational field, algebraic number field, field extension, algebraic number, inner product, linear processing of real vector-valued signals}
\end{IEEEkeywords}


\section{Introduction}\label{sec1}

Real numbers  form a field $\mathbf{R}$, i.e., they have arithmetics of 
addition, subtraction, multiplication, and division, and the multiplication
of two real numbers is commutative. This leads to the convenience in solving systems of 
linear equations of real coefficients, which has a fundamental 
importance 
in many science and engineering applications. 
A natural question is what happens to vectors of real values, i.e., real
vectors. When the dimension of real vectors is $2$, they also form a 
field that 
is the complex number field, by simply introducing the imaginary unit 
$\mathbf{i}=\sqrt{-1}$ as follows. For a two dimensional real vector $[x, y]$, let $z=x+ \mathbf{i} y$. 
Then, all $z$ form the complex number field $\mathbf{C}$ and  all the arithmetics 
of real vectors $[x, y]$ correspond to that of complex numbers $z$. 

However, for real vectors of dimension $3$ or higher, they do not form a field,
although all $4$ dimensional real vectors form the domain of quaternionic numbers
and all $8$ dimensional real vectors form the domain of octonionic numbers
\cite{ref10}. 
There exist arithmetics for $4$ dimensional real vectors and also for 
$8$ dimensional real vectors but their multiplications are not 
commutative.
This non-cummtativity may  cause a lot of inconvenience for the arithmetics of 
real vectors. 

Although higher dimensional real vectors do not form a field, if the real field is
replaced by a smaller subfield or a finite field,
the corresponding vectors may form a field.
For example, assume all the components of vectors of a fixed dimension $m$ 
take values in a finite field $\mathbf{E}$. Let $\alpha$ be a root of 
a primitive polynomial $p(x)$ of degree $m$ over $\mathbf{E}$
\cite{ref0, ref1}. 
Then, all the $m$ dimensional vectors of components taking values 
in $\mathbf{E}$ form another finite field $\mathbf{F}$ that is 
an extension of field $\mathbf{E}$ of degree $[\mathbf{F}:\mathbf{E}]=m$. 
This plays the fundamental
role in Reed-Solomon (RS) codes and BCH codes in error correction coding \cite{ref2}.
The rationale is explained in more details in Section \ref{sec5}. 

Another example is rational vectors, i.e., vectors with all components 
taking  
values in  the rational field
$\mathbf{Q}$. For $m$ dimensional rational vectors, let $\alpha$
have its minimal polynomial of degree $m$ over $\mathbf{Q}$.
Then, all the $m$ dimensional rational vectors form a field 
$\mathbf{Q}(\alpha)$ that is an algebraic number field and is 
an extension of $\mathbf{Q}$ of degree $m$. 
Therefore, all the $m$ dimensional rational vectors have the
$4$ arithmetics as real numbers 
and the multiplication of any two $m$ dimensional rational vectors is 
commutative. 

Since real values can be approximated by rational values to an arbitrary 
precision, in this paper we propose to use rational vectors to approximate
real vectors in their arithmetics within an $\epsilon$ range, called
$\epsilon$-arithmetics. We also define the complex conjugate of a real vector
and then define  inner product and linear convolution of two 
real vectors and two real vector sequences
(called vector-valued signals).
Thus, the proposed  $\epsilon$
arithmetics provide  similar
linear processing for general real vector-valued
signals as that for complex-valued signals.
This will  broaden the linear processing of the conventional real or complex-valued
signals, such as linear filtering,
ARMA modeling, and  least squares fitting to a set of data. 

Note that algebraic number fields have been applied in signal processing,
mostly in fast algorithms \cite{ref3, ref4},
communications, such as coding \cite{ref5}
and space-time coding \cite{ref6, ref7, ref8},
and cryptography, such as lattice based cryptography \cite{ref9}. However,
all these existing applications are different from what is proposed
in this paper 
for the arithmetics
  of addition, subtraction, multiplication, and division (in what follows, 
arithmetics always mean these four operations) of real vectors, 
which are used to linearly process real vector-valued
signals of any fixed vector size with real vector-valued coefficients
of the same vector size, similar to the conventional
real or complex scalar-valued signals. 

This paper is organized as follows. In Section \ref{sec2}, we define 
$\epsilon$-arithmetics for real vectors.
In Section \ref{sec3}, we define complex conjugate of a real
vector and  inner products for two 
real vectors and two real vector sequences.
In Section \ref{sec4}, we define convolution and linear filtering for finite length
real vector-valued signals. 
 In Section \ref{sec5}, 
we have some discussions. 
In Section \ref{sec5}, we conclude this paper.

\section{$\epsilon$-Arithmetics}\label{sec2}

We first briefly introduce an algebraic number field, an extension of 
the rational field  $\mathbf{Q}$. 
Let $\alpha$ be an algebraic number with its minimal polynomial $p(x)$ of degree $m$ over the rational field 
$\mathbf{Q}$, i.e., $p(x)$ is the polynomial of lowest degree
with coefficients in the rational field $\mathbf{Q}$ such that
$\alpha$ is a root of the polynomial. Then, 
\begin{equation}\label{1}
\mathbf{Q}(\alpha)=\{ \left. q_1+q_2\alpha+\cdots + q_m \alpha^{m-1} \right| q_i\in \mathbf{Q}, 1\leq i\leq m   \}
\end{equation}
is an algeraic number field and an extension of $\mathbf{Q}$ of degree $m$.
Thus, all elements in $\mathbf{Q}(\alpha)$ have the $4$ conventional 
arithmetics and the multiplication is commutative. 
Let $\mathbf{Q}[x]$ denote the ring of all polynomials over $\mathbf{Q}$, i.e., polynomials
with rational coefficients. Then, the
field $\mathbf{Q}(\alpha)$ is isomorphic to
$\mathbf{Q}[x]/(p(x))$, the field of
all polynomials under modulo $p(x)$ operation,
 i.e., $\mathbf{Q}(\alpha)\cong \mathbf{Q}[x]/(p(x))$.

We next introduce the arithmetics for rational vectors.
For any $m$ dimensional rational
vector $\mathbf{q}=[q_1,q_2,\cdots, q_m]\in \mathbf{Q}^m$, we map
$\mathbf{q}$ to the element $q_1+q_2\alpha+\cdots +q_m \alpha^{m-1}$ in
$\mathbf{Q}(\alpha)$ and it is clear that this mapping is one-to-one and onto.
We define the arithmetics of these $m$ dimensional rational vectors
as that of their mapped elements in the algebraic number field 
$\mathbf{Q}(\alpha)$:
$$
[q_{11},q_{12},\cdots,q_{1m}]\circ [q_{21},q_{22},\cdots,q_{2m}]
$$
\begin{equation}\label{2}
\stackrel{\Delta}{=}\sum_{i=1}^m q_{1i}\alpha^{i-1} \circ \sum_{i=1}^m q_{2i}\alpha^{i-1}
\end{equation}
where ${\bf q}_j=[q_{j1},q_{j2},\cdots, q_{jm}]\in \mathbf{Q}^m$ for $j=1,2$, and 
$\circ$ is an arithmetic operation, i.e., one of addition, subtraction,
 multiplication and division. 
Since $\mathbf{Q}(\alpha)$ is a field, the right hand side of (\ref{2}) 
is also an element in  $\mathbf{Q}(\alpha)$,  and thus 
has an expression $q_1+q_2\alpha+\cdots +q_m \alpha^{m-1}$ for 
some $q_i\in \mathbf{Q}$, $1\leq i\leq m$. Then, 
$$
[q_{11},q_{12},\cdots,q_{1m}]\circ [q_{21},q_{22},\cdots,q_{2m}]
$$
\begin{equation}\label{201}
=[q_1,q_2,\cdots,q_m]\in \mathbf{Q}^m.
\end{equation}

With $\mathbf{Q}(\alpha)\cong \mathbf{Q}[x]/(p(x))$, the addition, subtraction,
multiplication of two rational vectors of dimension $m$ are easy to implement.
To do division, we only need to know how to implement the inverse of a non-zero
rational vector $[q_1,q_2,\cdots, q_m]\neq 0 \in \mathbf{Q}^m$. 
Let $q(\alpha)=q_1+q_2\alpha+\cdots+q_m\alpha^{m-1} \neq 0 \in 
\mathbf{Q}(\alpha)$. 
Let its inverse be $\bar{q}(\alpha)=\bar{q}_1+\bar{q}_2\alpha
+\cdots +\bar{q}_m \alpha^{m-1}$, i.e., 
\begin{equation}\label{202}
 \bar{q}(\alpha)  q(\alpha) =1 \mod p(\alpha).
\end{equation}
Since $p(x)$ is a minimal polynomial over $\mathbf{Q}$, 
$q(x)$ and $p(x)$ are co-prime over $\mathbf{Q}$. 
From B\'{e}zout's identity and using Euclidean algorithm, one can
find\footnote{It can be done by using Matlab easily.}
two polynomials of rational coefficients $u(x), v(x)\in \mathbf{Q}[x]$
 such that
\begin{equation}\label{203}
u(\alpha) q(\alpha)+ v(\alpha) p(\alpha)=1,
\end{equation}
which implies $u(\alpha)q(\alpha)=1 \mod p(\alpha)$, and thus
we have $\bar{q}(\alpha)=(q(\alpha))^{-1}=u(\alpha) \mod p(\alpha)$ in (\ref{202}). 
Then, $([q_1,q_2,\cdots,q_m])^{-1}=[\bar{q}_1,\bar{q}_2,\cdots, \bar{q}_m]
\in \mathbf{Q}^m$.
A detailed Matlab code to compute the inverse of a real vector
can be found in the Appendix. 
With the inverse calculation of a non-zero $m$ dimensional 
rational vector, the divisions of non-zero $m$ dimensional rational vectors 
follow immediately.

Note that, since the algebraic number field 
$\mathbf{Q}(\alpha)$ is a subfield of the complex number field $\mathbf{C}$,
the arithmetic, $\circ$, in (\ref{2}) on the algebraic number 
field $\mathbf{Q}(\alpha)$
is also the same as that on the complex number field, i.e., the conventional 
$+,-,\times, \div$ for complex numbers.

From the above definition of the arithmetics for rational vectors, one can see
that different algebraic numbers $\alpha$ define different arithmetics for 
the same rational vectors. Let us see two examples. Let $m=4$, 
$\alpha_1=\sqrt{2}+\sqrt{3}$, and its minimal polynomial is 
$p_1(x)=x^4-10 x^2+1$, and let $\alpha_2=\exp(2\pi \mathbf{i}/5)$, and the  
minimal polynomial is $p_2(x)=x^4+x^3+x^2+x+1$. In fact, $\alpha_2$ is 
a cyclotomic number and its generated algebraic number field is a cyclotomic 
field \cite{ref1}. 

For simplicity, consider two $4$ dimensional rational vectors,  
$[1,1,1,1]$ and $[1,1,-1,-1]$, and their multiplications following 
(\ref{2}) and  the  two algebraic numbers $\alpha_1$ and 
$\alpha_2$. For the multiplication of the two rational vectors
with $\alpha_1$, 
 following (\ref{2}) and $\alpha_1^4=10\alpha_1^2-1$ due to $p_1(\alpha_1)=0$, 
 we have  
\begin{eqnarray*}
 & [1,1,1,1]\cdot [1,1,-1,-1]\\
 = & 1  +\alpha_1  -\alpha_1^2  -\alpha_1^3\\
   & +1 + \alpha_1 -9 \alpha_1^2 -\alpha_1^3\\
   & +1 +\alpha_1 -9 \alpha_1^2 -9 \alpha_1^3\\
   & +9 +\alpha_1-89 \alpha_1^2-9 \alpha_1^3\\
 = & 10+4\alpha_1 -108 \alpha_1^2 -20 \alpha_1^3\\
 = & [12, 4, -108, -20].
\end{eqnarray*}
For the multiplication of the two rational vectors
with $\alpha_2$, following (\ref{2}) and $\alpha_2^4=-\alpha_2^3-\alpha_2^2-\alpha_2-1$ due to $p_2(\alpha_2)=0$, we have 
\begin{eqnarray*}
 & [1,1,1,1]\cdot [1,1,-1,-1]\\
 = & 1  +\alpha_2  -\alpha_2^2  -\alpha_2^3\\
   & +1 + 2\alpha_2 + 2 \alpha_2^2 \\
   &  +\alpha_2 +2 \alpha_2^2 + 2 \alpha_2^3\\
   & -2 -2\alpha_2- \alpha_2^2\\
 = & 2\alpha_2 +2 \alpha_2^2 +\alpha_2^3\\
 = & [0,2,2,1].
\end{eqnarray*}
One can see that the above two multiplication results of the same  two rational vectors are much different.

We now introduce $\epsilon$-arithmetics for real vectors for a fixed 
algebraic number $\alpha$ with its minimal  polynomial of degree $m$ 
over $\mathbf{Q}$. Consider the 
$m$ dimensional real vector space $\mathbf{R}^m$ for $m\geq 3$. 
Let $\epsilon>0$ be an arbitrary small positive number. Let 
$$
{\bf r}_1=[r_{11},r_{12},\cdots, r_{1m}], {\bf r}_2 =
[r_{21},r_{22},\cdots, r_{2m}]\in \mathbf{R}^m
$$
 be two arbitrary 
$m$ dimensional real vectors. Their $\epsilon$-arithmetics are defined
as follows. 

Find two $m$ dimensional rational vectors
${\bf q}_1, {\bf q}_2\in \mathbf{Q}^m$ in the 
$\epsilon$ ranges of ${\bf r}_1, {\bf r}_2$:
\begin{equation}\label{3}
\| {\bf r}_j-{\bf q}_j\|<\epsilon, \,\,\, j=1,2,
\end{equation}
where $\| \cdot \|$ is a norm for $m$ dimensional vectors, such as
 $l_2$ or $l_{\infty}$ norm,  and 
if any ${\bf r}_j$ is rational, then ${\bf q}_j={\bf r}_j$, $j=1$ or/and $2$. 

The $\epsilon$-arithmetic operation for two $m$ dimensional 
vectors ${\bf r}_1$ and ${\bf r}_2$  is defined as
\begin{equation}\label{4}
{\bf r}_1 \circ {\bf r}_2
\stackrel{\Delta}{=} {\bf q}_1\circ {\bf q}_2,
\end{equation}
where $\circ$ is an arithmetic operation, such as 
$+,-,\times, \div$, and ${\bf q}_1\circ {\bf q}_2$ is defined 
in (\ref{2})-(\ref{201}).

From the above definition of $\epsilon$-arithmetics, clearly the 
$\epsilon$-arithmetic result of two real vectors ${\bf r}_j$, $j=1,2$,  is 
not unique, even for a fixed algebraic number $\alpha$ in (\ref{2}). 
This is because a rational vector ${\bf q}_j$ in (\ref{3}) 
in the $\epsilon$ range of the real vector ${\bf r}_j$ is not unique.
In fact, the above $\epsilon$-arithmetics in (\ref{4}) can be
defined as a set-valued  
mappings, where ${\bf r}_1 \circ {\bf r}_2$ is  equal to a set of  
$ {\bf q}_1\circ {\bf q}_2$ in (\ref{4}) for non-empty sets of  
${\bf q}_1$ and ${\bf q}_2$ in the $\epsilon$ ranges of
${\bf r}_1$ and ${\bf r}_2$ in (\ref{3}), respectively.
Although this is the case, since rational numbers are dense 
in the real field $\mathbf{R}$, 
 this $\epsilon$ can be made arbitrarily small, such as 
a computer numerical error level. In this case, within
the numerical precision range, the error (or difference) 
 in the  $\epsilon$-arithmetics
from different rational vector approximations in the 
$\epsilon$ ranges of two real vectors is negligible, or just 
the computer numerical error.
Note that this may be similarly  done as  in \cite{ref4} 
by using arbitrary large integers in getting 
rational vectors with a desired precision.

We know that all real values  stored in
computers must be rational. Therefore, in terms of practical  
computations on computers, the $\epsilon$-arithmetics for 
real vectors can be made the same as the arithmethics for rational vectors. 
Thus, for notational convenience and without confusion in understanding,
for two real vectors ${\bf r}_j=[r_{j1},r_{j2},\cdots, r_{jm}]\in \mathbf{R}^m$, $j=1,2$, their
arithmetic $\circ$ is also written as
\begin{equation}\label{2000}
{\bf r}_1\circ {\bf r}_2 
=\sum_{i=1}^m r_{1i}\alpha^{i-1} \circ \sum_{i=1}^m r_{2i}\alpha^{i-1},
\end{equation}
and 
use the following abusively for ${\bf r}=[r_1,r_2,\cdots, r_m]\in \mathbf{R}^m$:
\begin{equation}\label{2001}
{\bf r}=\sum_{i=1}^m r_i \alpha^{i-1}.
\end{equation}

Note that because the $\epsilon$-arithmetics for real vectors
using approximated rational vectors are defined 
in (\ref{2}) with only a fixed $m$ finite
opeartions, they are robust to the
approximation errors.
For a real vector that is in an $\epsilon$
range of $0$, then it is treated as $0$. Otherwise, its division is also
robust to an approximation error.
This means that when $\epsilon$ is small enough,
the differences of  $\epsilon$-arithmetic operations 
using different rational vector 
approximations of real vectors is negligible in practical calculations
of $\epsilon$-arithmetics.
A detailed Matlab code to compute the product  of two real vectors 
can be found in the Appendix.


With the above $\epsilon$-arithmetics for real vectors, one is able
to systematically solve systems of linear equations over real vectors: 
\begin{equation}\label{5}
\sum_{j=1}^J {\bf a}_{lj} \cdot {\bf x}_j ={\bf b}_l, \,\,\, l=1,2,...,L,
\end{equation}
where ${\bf a}_{lj}$, ${\bf b}_l$ are known real vectors 
of dimension $m$ in $\mathbf{R}^m$ and ${\bf x}_j$ are unknown 
real vectors of dimension $m$ to solve. It may have  applications in,
for example, the least squares fitting to a set of data, which may be broader than
the conventional  least squares fitting. 

As a remark, for convenience, the above real vector ${\bf r}\in \mathbf{R}^m$
always corresponds to the algebraic number $\alpha$ with its minimal  
polynomial $p(x) \in \mathbf{Q}[x]$ of degree $m$,
unless otherwise specified.

\section{Complex Conjugate and  Inner Product of Real Vectors}\label{sec3}

In this section, we define complex conjugate and inner product for
real vectors when the algebraic number $\alpha$ is a real number or cyclotomic
number $\exp(2\pi\mathbf{i}/p)$ for a prime number $p$ with $p>2$.

When the algebraic number  $\alpha$ is real, the complex conjugate 
of a real vector ${\bf r}$ is defined as itself, i.e., ${\bf r}^* ={\bf r}$.
When the algebraic number $\alpha$ is cyclotomic
 $\exp(2\pi\mathbf{i}/p)$ for a prime number $p$ with $p>2$, 
the complex conjugate of $\alpha$ is 
$$
\alpha^* =\exp(-2\pi\mathbf{i}/p)=\alpha^{-1}=\exp(2\pi\mathbf{i}(p-1)/p)
$$
\begin{equation}\label{3.1}
  =\alpha^{(p-1)} =-\sum_{i=0}^{p-2}\alpha^{i},
\end{equation}
since, in this case, the minimal polynomial is $1+\alpha+\cdots +\alpha^{p-1}$
\cite{ref1}. Then, for a real vector ${\bf r}=[r_1,r_2,\cdots,r_m]\in \mathbf{R}^m$, its complex conjugate
${\bf r}^*$ is defined as
$$
{\bf r}^* \stackrel{\Delta}{=}\sum_{i=1}^m r_i (\alpha^*)^{i-1} 
= \sum_{i=1}^m r_i \alpha^{-(i-1)}
$$
\begin{equation}\label{3.2}
= \sum_{i=1}^m r_i \alpha^{(i-1)m}
 =\sum_{i=1}^m r_i \left( -\sum_{k=0}^{m-1}\alpha^k\right)^{i-1},
\end{equation}
where $m=p-1$. 
Clearly $({\bf r}^*)^*={\bf r}$. Let us see a simple
example when $p=3$. Then, $m=2$ and 
$$
{\bf r}^* = r_1 -r_2(1+\alpha)=r_1-r_2 -r_2\alpha
     =[r_1-r_2, -r_2].
$$
This is just for an illustration,  since for $m=2$ dimensional real vectors,
one may simply use the complex field $\mathbf{C}$ for the arithmetics.

For two real vectors ${\bf r}_{j}=[r_{j1},r_{j2},\cdots,r_{jm}]\in \mathbf{R}^m$, $j=1,2$, their inner product is defined as
\begin{equation}\label{3.3}
\langle {\bf r}_1, {\bf r}_2\rangle
={\bf r}_1 \cdot {\bf r}_2^*,
\end{equation}
where ${\bf r}_2^*$ is the complex conjugate of ${\bf r}_2$.
Clearly, we have 
$\langle {\bf r}_1, {\bf r}_2 \rangle
=(\langle {\bf r}_2, {\bf r}_1 \rangle)^*$.

For two real vector sequences 
$\bar{{\bf r}}_j=\{{\bf r}_{j,l}\}_{1\leq l\leq L}$ 
of dimension $m$ and length $L$ for $j=1,2$, their inner product is 
defined as: 
\begin{equation}\label{3.4}
\langle \bar{{\bf r}}_1, \bar{{\bf r}}_2 \rangle  \stackrel{\Delta}{=} 
\sum_{l=1}^L {\bf r}_{1,l} \cdot {\bf r}_{2,l}^*,
\end{equation}
where ${\bf r}_{2,l}^*$ is the complex conjugate of 
${\bf r}_{2,l}$.

Two real vectors ${\bf r}_i$, $i=1,2$, are called orthogonal
if their inner product defined in (\ref{3.3}) is 0.
Two real vector sequences $\bar{\bf r}_i$, $i=1,2$, of length $L$
are called orthogonal, if their inner product defined in (\ref{3.4}) is $0$.

For the inner product of two real vectors defined in (\ref{3.3}), 
it can be expanded as
\begin{equation}\label{3.5}
\langle {\bf r}_1, {\bf r}_2\rangle
=\sum_{i=1}^m r_{1i}\alpha^{i-1} \sum_{i=1}^m r_{2i} (\alpha^*)^{i-1}.
\end{equation}
Let us see an example of $4$ dimensional real vectors in $\mathbf{R}^4$
with $\alpha=\sqrt{2}+\sqrt{3}$ and its minimal polynomial 
$p(x)=x^4-10x^2+1$. In this case, by some algebra the inner product of 
${\bf r}_{j}=[r_{j1},r_{j2},r_{j3},r_{j4}]\in \mathbf{R}^4$, $j=1,2$, 
is
\begin{eqnarray}
\langle {\bf r}_1, {\bf r}_2\rangle  & = &
        [r_{11}r_{21}-r_{12}r_{24} -r_{13}r_{23}\nonumber \\
     & &    \hspace{0.3in} -r_{14}(r_{22}+10 r_{24}),  \nonumber \\
      & & r_{11}r_{22}+r_{12}r_{21}-r_{13}r_{24}-r_{14}r_{23},  \nonumber \\
     & & r_{11}r_{23}+r_{12}(r_{22}+10r_{24})+r_{13}(r_{21}+10r_{23})  \nonumber \\
       & &  \hspace{0.3in} +r_{14}(10r_{22}+99r_{24}), \nonumber \\
    & & r_{11}r_{24}+r_{12}r_{23}+r_{13}(r_{22}+10r_{24}) \nonumber \\
       & & \hspace{0.3in} +r_{14}(r_{21}+10r_{23})]. \label{3.6}
  \end{eqnarray}

When ${\bf r}_2={\bf r}_1={\bf r}=[r_1,r_2,\cdots, r_m]\in \mathbf{R}^m$,
 the squared norm of ${\bf r}$ becomes 
\begin{eqnarray*}
\langle {\bf r}, {\bf r} \rangle   & = & 
[r_1^2-r_3^2-10r_4^2-2r_2r_4,\\
 & &  2r_1r_2-2r_3r_4, \\
  & & r_2^2+10r_3^2+99r_4^2+2r_1r_3+20r_2r_4, \\
 & & 2r_1r_4+2r_2r_3+20r_3r_4].
\end{eqnarray*}

From (\ref{3.5}), the right hand side is the product of two polynomials and 
the coefficients of the product of two polynomials would come from the 
convolution of the two vectors in general, i.e., the inner product 
of the real vectors would be their convolution. 
From (\ref{3.6}), one can see that it is clearly not true here, which is due 
to the modulo $p(x)=x^4-10x^2+1$ operation, i.e., 
$\alpha^4=10\alpha^2-1$, in the product of two polynomials. 

If the product of two polynomials in the right hand side of (\ref{3.5})
was under the modulo $p(x)=x^4-1$, i.e., $\alpha^4=1$, then, 
the inner product $\langle {\bf r}_1, {\bf r}_2\rangle$ would be the 
circular convolution of the two vectors:
\begin{eqnarray*}
\langle {\bf r}_1, {\bf r}_2\rangle & = & 
   (r_{11}+r_{12}\alpha+r_{13}\alpha^2+r_{14}\alpha^3)\\
   & & \cdot (r_{21}+r_{22}\alpha^{-1}+r_{23}\alpha^{-2}+r_{24}\alpha^{-3})\\
   & = &  [r_{11}r_{21}+r_{12}r_{22}+r_{13}r_{23}+r_{14}r_{24},\\
& &  r_{11}r_{24}+r_{12}r_{21}+r_{13}r_{22}+r_{14}r_{23}, \\
& &  r_{11}r_{23}+r_{12}r_{24}+r_{13}r_{21}+r_{14}r_{22}, \\
& &  r_{11}r_{22}+r_{12}r_{23}+r_{13}r_{24}+r_{14}r_{21}].
\end{eqnarray*}
This would correspond to the product of the $4$-point DFTs
of the two vectors of size $4$.
It means that the four point
 evenly spaced samplings in the frequency domain determines the vector 
of length $4$. 
However, $p(x)=x^m-1$ cannot be a minimal polynomial 
for any positive integer $m>1$ and thus the inner product 
of two real vectors cannot be a circular convolution in general.

As another comparison, we consider the algebraic number $\alpha=\alpha_2$ in Section \ref{sec2}, i.e., 
$\alpha=\exp(2\pi\mathbf{i}/5)$. Then,   $\alpha^5=1$ and 
$\alpha^4=-(\alpha^3+\alpha^2+\alpha+1)$. In this case, by some algebra we have
\begin{eqnarray*}
\langle {\bf r}_1, {\bf r}_2\rangle & = &
   (r_{11}+r_{12}\alpha+r_{13}\alpha^2+r_{14}\alpha^3)\\
   & & \cdot (r_{21}+r_{22}\alpha^{-1}+r_{23}\alpha^{-2}+r_{24}\alpha^{-3})\\
   & = &
        [r_{11}(r_{21}-r_{22})+r_{12}(r_{22}-r_{23})\\
& &    \hspace{0.3in} +r_{13}(r_{23}-r_{24})+r_{14}r_{24},\\
          & &  -r_{11}r_{22}+r_{12}(r_{21}-r_{23})\\
& &    \hspace{0.3in} +r_{13}(r_{22}-r_{24})+r_{14}r_{23},\\
          & &  r_{11}(r_{24}-r_{22})-r_{12}r_{23}\\
          & & \hspace{0.3in} +r_{13}(r_{21}-r_{24})+r_{14}r_{22},\\
          & &  r_{11}(r_{23}-r_{22})+r_{12}(r_{24}-r_{23})\\
          & & \hspace{0.3in} +r_{13}(r_{21}-r_{24})+r_{14}r_{21}].
\end{eqnarray*}

For two general $m$ dimensional
 real vectors ${\bf r}_1$ and ${\bf r}_2$
 with a general algebraic number $\alpha$
with its minimal polynomial 
$p(x)=p_1+p_2x+\cdots +p_mx^{m-1}+x^m$
for some $p_i\in \mathbf{Q}$, $1\leq i\leq m$, 
 if $\alpha^*\in \mathbf{Q}(\alpha)$ as 
\begin{equation}\label{3.7}
\alpha^* =\sum_{i=1}^m a_i \alpha^{i-1},
\end{equation}
where $a_i\in \mathbf{Q}$, $1\leq i\leq m$, 
then their inner product can be similarly defined as (\ref{3.5}):
\begin{eqnarray*}
\langle {\bf r}_1, {\bf r}_2\rangle 
 & = & \sum_{i=1}^m r_{1i}\alpha^{i-1} \sum_{i=1}^m r_{2i} (\alpha^*)^{i-1}\\
 & = & \sum_{i=1}^m r_{1i}\alpha^{i-1} \sum_{i=1}^m r_{2i} 
 \left( \sum_{k=1}^m a_k \alpha^{k-1}\right)^{i-1}\\
 & = & \sum_{i=1}^m r_i \alpha^{i-1}\\
 & = & [r_1,r_2,\cdots,r_m],
\end{eqnarray*}
where, after the polynomial  expansions,  each $r_i$, $1\leq i\leq m$, 
 is a linear function of $r_{1k}r_{2n}$,
$1\leq k, n\leq m$, with  coefficients in $\mathbf{Q}$  that are
some functions of $p_{l_1}, a_{l_2}$, $1\leq l_1,l_2\leq m$.

For the inner product of two real vectors, from (\ref{3.5}), one can see that the right hand side is a product of two
complex numbers. Thus, $\langle {\bf r}_1, {\bf r}_2\rangle =0$ if and only if
either ${\bf r}_1=0$ or ${\bf r}_2=0$. Then, 
$\langle {\bf r}, {\bf r}\rangle =0$
if and only if ${\bf r}=0$. And two real vectors are orthogonal if and only if
one of them is $0$. It is also not hard to see
that for a real vector sequence $\bar{\bf r}$, $\langle \bar{\bf r}, \bar{\bf r}\rangle =0$ if and only if $\bar{\bf r}=0$, i.e., the $0$ sequence.

For two rational vector sequences $\bar{\bf q}_i$, $i=1,2$, of length $L$,
from (\ref{3.4}), (\ref{3.3}), and (\ref{3.5}), they are orthogonal
if and only if their corresponding sequences of complex numbers
are orthogonal. This implies that there are at most  $L$ many
orthogonal rational vector sequences of length $L$.
Furthermore, from Gram-Schmidt orthogonalization procedure,
it is not
hard to see that there  exist $L$ many orthogonal real 
vector sequences of length $L$.

\section{Convolution and Linear Filtering of Real Vector-Valued Signals}\label{sec4}

With the above inner product for real vectors, it is easy to define
a convolution of two real vector sequences (or called real vector-valued
signals).
Let ${\bf r}_j(n)$, $j=1,2$, be two real vector-valued signals of finite length, i.e.,
for each integer $n$ in a finite range,
${\bf r}_j(n)$ is a real vector defined as before
and $0$ for other $n$ for $j=1,2$. Their convolution is defined as
\begin{equation}\label{4.1}
 {\bf r}_1(n) \ast {\bf r}_2 (n) = \sum_k \langle {\bf r}_1 (k), {\bf r}_2(n-k)\rangle = \sum_k {\bf r}_1(k)\cdot {\bf r}_2^*(n-k).
\end{equation}
It is clear that when the vector size is $2$ and the primitive element
$\alpha= \mathbf{i}$, the above convolution coincides with the convolution of
two complex-valued signals. For a general vector size, since
the $\epsilon$-arithmetic
operations of real vectors, i.e., approximated rational vectors,
are all commutative,
all the properties of convolutions for complex-valued signals hold for
general real vector-valued signals.
Thus, if one real vector-valued signal is treated as a filter impulse
response, say ${\bf h}(n)={\bf r}_1(n)$,
and the other is an input signal, ${\bf s}(n)={\bf r}_2(n)$,
the above convolution is
the linear filtering of a real vector-valued input signal  ${\bf s}(n)$
to a real vector-valued system ${\bf h}(n)$. 

Note that the above definition of convolution only applies to
finite length real vector-valued signals. This is because in  the
$\epsilon$-arithmetics for real vectors, it uses rational approximations
within an $\epsilon$ range. When there are infinitely many 
$\epsilon$-arithmetics in a summation, the approximation error
may blow up, no matter how small an $\epsilon$ is.
Fortunately, in practice all signals in computations have finite length
and thus the above convolution for finite length real vector-valued
signals is sufficient.

\section{Some Discussions}\label{sec5}

After the earlier definitions of arithmetics, inner product, convolution,
and linear filtering for real vectors and vector-valued signals,
one might ask whether there are any applications.
To explain some of their potential applications, 
we first explain how the arithmetics of vectors of finite many
elements (in finite fields) are used in discovering better error correction codes. 

In error correction codes, it is well-known that 
in order to have computationally  efficient
 encoding and decoding, one usually uses linear 
codes \cite{ref2}. 
Before using  vector arithmetics (or finite fields), linear block
 error correction codes were binary linear block codes where generator  
matrices are binary, i.e., matrix entries are either $0$ or $1$. This 
may strongly limit the 
possibilities and
choices of good generator matrices. In order to have more choices
for good generator matrices, it is critical to expand the choices of binary 
elements of the entries in a generator matrix. One common way to do so is 
to generalize binary scalar values $0$ and $1$ to vectors of binary values, 
i.e., binary vectors. Then, the question is how to do the arithmetics for
binary vectors. In error correction coding, to correct errors it is 
important to be able to solve systems of linear equations over, in this case,
 binary vectors. To do so, it is important to have all arithmetics for binary
vectors. As what was mentioned in Introduction, 
the field extension over the binary field $\{0,1\}$ provides such a property, 
i.e., it provides all the arithmetics for vectors of finite many elements. 
This is the foundation on how RS codes and BCH codes are constructed \cite{ref2}. 
Without 
 finite fields (or arithmetics of vectors of finite many elements), 
it would not exist RS codes or BCH codes that have been widely used 
in our daily life electronics.

Let us see an
example for the above rationale. Consider a binary linear block code
with a generator matrix of size $3\times 2$, i.e., $2$ input binary
symbols produce $3$ output binary symbols. In this case, there are total
$2^6=64$ possible generator matrices to choose. Now we replace every element
in a $3\times 2$ binary generator matrix by a binary vector of size $m=4$,
i.e., an element in Galois field GF($2^4$). As an example, consider the following binary
$3\times 2$ generator matrix:
$$
\left[ \begin{array}{cc} 1 & 0 \\
    0 & 1\\
    1 & 1
       \end{array}
    \right].
$$
We now replace $0$ by $[0,0,0,0]$ and $1$ by $[1,0,0,0]$ in the above binary
matrix and  obtain
\begin{equation}\label{5.1}
\left[ \begin{array}{cc} $[1,0,0,0]$ & $[0,0,0,0]$ \\
    $[0,0,0,0]$ & $[1,0,0,0]$\\
    $[1,0,0,0]$ & $[1,0,0,0]$
    \end{array}
    \right].
\end{equation}
In this case, the encoding is to multiply this $3\times 2$ matrix with
a $2\times 1$ information symbol vector of two binary vectors of size $4$
from the right. 

Note that a $2\times 1$ information symbol vector
of two binary vectors of size $4$ 
is the same as
a $8\times 1$ binary vector as a whole.
Thus, if the above encoding over GF($2^4$) needs to be compared with
a linear binary encoding, for example, one might ask whether it corresponds
to a valid binary linear block code? If it does,
the binary generator matrix would be
$$
\left[ \begin{array}{cccccccc}
    1 & 0 & 0 & 0 & 0 & 0 & 0 & 0\\
    0 & 0 & 0 & 0 & 1 & 0 & 0 & 0\\
    1 & 0 & 0 & 0 & 1 & 0 & 0 & 0
  \end{array}
  \right],
$$
by lining up the binary components in the binary vectors of size $4$
in each row in the matrix in (\ref{5.1}) into a new row. 
One can then clearly see that this binary matrix is $3\times 8$ and not even
invertible from the left,
i.e., it cannot be a valid generator matrix for a decodable binary
linear block code. In other words, the linear encoding
over finite field GF($2^4$) does not correspond to a binary linear block code.
There are much more valid linear block codes over a larger size
finite field than those over the binary field.

The same idea as the binary vectors can be applied to real vectors.
The vectorized linear processing
broadens the scalar linear processing and provides much more degrees of freedom
and convenience, while the linear arithmetics of real vectors
provide
the convenience in finding solutions in practical applications.
In addition to the linear filtering proposed in Section \ref{sec4}, 
another  application is the least squares fitting to a set of data.
With the $\epsilon$-arithmetics defined for real vectors in this paper,
compared to the conventional least squares  scalar fitting,
the least squares vector
fitting is more flexible and therefore,  better
fitting performance can be expected.
A similar
application is the linear autoregressive and moving average (ARMA)
model fitting in time series.
In summary, the $\epsilon$-arithmetics introduced in this paper
will open the door to enlarge the pool of all the existing
linear processing techniques of scalar-valued
signals.

Another potential application is in image processing, such as image
compression. Usually, image pixels are correlated in two dimensions, while
the conventional vector based signal processing techniques, such as vector
transforms (VT) first proposed and applied in image compression
by Li in \cite{vt1, vt2} and investigated later
in \cite{vt3, vt66,  vt4, vt5, vt6, vt7, vt8}, vector Karhunen-Lo\`{e}ve
transform (KLT) \cite{vt7}, and vector-valued wavelets \cite{vt9},
may only decorrelate signals in one dimension.
And they are usually applied row-wise and coloumn-wise separately. 
An extended VT called average optimal VT was proposed in \cite{vt8}
for a two dimensional image by doing average in
one dimension first and doing VT
in the other dimension second to the blocks of the image.

With the arithmetics for real vectors proposed in this paper,
for a block of an image,
we may treat it as a real vector-valued signal, i.e.,  
consider its first dimenion as a vector value ${\bf r}$   and
the second  dimension as the vector $[{\bf r}_1, {\bf r}_2, \cdots, {\bf r}_L]$
of the vector values ${\bf r}_i$ for $i=1,2,...,L$.
Then, we may apply a vector-valued signal processing technique,
such as VT, vector KLT, or vector-valued wavelets, to this vector.
This proposed processing might be possible to decorrelate a two dimensional
image along both dimensions simultaneously. 
We believe that it will provide a significantly  new technique for image
processing. 
Note that what is proposed in this paper 
is different from matrix KLT and matrix-valued wavelets for processing
matrix-valued signals directly in \cite{vt10}.

As a remark,
an issue about the $\epsilon$-arithmetics for real vectors defined in this
paper is the choice of an algebraic number $\alpha$ that determines the 
$\epsilon$-arithmetics for real vectors. As mentioned earlier, different 
 algebraic numbers $\alpha$ provide different 
$\epsilon$-arithmetics for real vectors and the difference may be large.
Then, the question is which algebraic number $\alpha$ is needed or better.
The answer to this question might depend on the application scenario.

\section{Concluding Remarks}\label{sec6}
It is known that real vectors of dimension higher than $2$ do not form a field. In other words, 
one cannot do arithmetics for real vectors of dimension higher than $2$ 
similar to that for real numbers. This  may limit the capability of 
solving systems of linear equations over real vectors of a finite dimension.
In this paper, we have proposed $\epsilon$-arithmetics for real vectors 
by using approximations of rational vectors that form algebraic number
fields. We have also 
defined the complex conjugate of a real vector, inner product, and convolutions
of two real vectors and two real vector sequences (or called vector-valued
signals). This may lead to
systematic linear processing for real vector-valued signals  with real vector-valued coefficients, 
such as linear filtering, least squares fitting,
and ARMA modeling to real vector-valued
signals.
The $\epsilon$-arithmetics proposed in this paper for real vectors
provide the same convenience as that for complex numbers in linear
processing, while they broaden the conventional
linear processing of scalar-valued signals.
It is believed that the study in this paper will open a door to
the signal processing community.

\section*{Appendix: Matlab Codes to Compute Real Vector Multiplication
  and Inverse}

Below are  Matlab function codes to compute the
product of real vectors ${\bf p}_1$ and ${\bf p}_2$, and the 
inverse ${\bf p}^{-1}$ of a real vector ${\bf p}$,
for a given minimal polynomial $q(x)$.
The vector of the coefficients of $q(x)$ is ${\bf q}$.
In all real vectors, i.e., all coefficient vectors of polynomials,
in the following Matlab codes, their components are in the decreasing order:
${\bf q}(1)=q_{m}$, ${\bf q}(2)=q_{m-1}$, ..., ${\bf q}(m)=q_1$, if $q(x)=\sum_{k=0}^{m-1}
q_{k+1} x^{k}$.  This is opposite to the descriptions in this paper, where
they are in the increasing order. In other words, the orders
of vector components in the paper and in the following Matlab codes are
the reverse of each other.

\subsection*{Real Vector Multiplication}

\begin{itemize}
\item[]  $p1$ and $p2$ are are two real vectors of dimension $m$ to multiply. 
\item[]  $p1$, $p2$, and $q$ are row coefficient vectors of polynomials
        as $p(x)$ and $q(x)$. 
\item[]  $q(x)$ is a minimal polynomial of degree $m-1$.
\item[]  In a polynomial, the order of $x$ is from high to low.
     For example, if $p=[2,1,3]$, then $p(x)=2x^2+x+3$. 
\item[] The output mvec is the product of $p1$ and $p2$. 
\end{itemize}

\vspace{0.3in}

function[mvec]=vectormultiply(p1,p2,q)

syms x;

psize=size(p1);

qsize=size(q);

mvec=0$\ast$p1;

pp1=x.\^{}[psize(2)-1:-1:0]$\ast$p1\'{};

pp2=x.\^{}[psize(2)-1:-1:0]$\ast$p2\'{};

qq=x.\^{}[qsize(2)-1:-1:0]$\ast$q\'{};

pp=pp1$\ast$pp2;

[Q,R]=quorem(pp,qq,x);

mv=sym2poly(R);

mvsize2=size(mv);

mvsize=mvsize2(2);

mvec(psize(2)-mvsize+1:1:psize(2))=mv;

\subsection*{Inverse of a Real Vector}

\begin{itemize}
\item[] $p$ is a real vector of dimension $m$ to have its inverse. 

\item[] $p$ and $q$ are row coefficient vectors of polynomials $p(x)$ and
  $q(x)$, respectively.

\item[]  $q(x)$ is a minimal polynomial of degree $m-1$.

\item[] In a polynomial, the order of $x$ is from high to low. For example, if
 $p=[2,1,3]$, then $p(x)=2x^2+x+3$. 

\item[]  The output ivec is the inverse of real vector $p$.
  \end{itemize}

\vspace{0.3in}

function[ivec]=vectorinverse(p,q)

syms x;

psize=size(p);

qsize=size(q);

ivec=0$\ast$p;

pp=x.\^{}[psize(2)-1:-1:0]$\ast$p\'{};

qq=x.\^{}[qsize(2)-1:-1:0]$\ast$q\'{};

[g,c,d]=gcd(pp,qq,x);

iv=sym2poly(c); 

ivsize2=size(iv);

ivsize=ivsize2(2);

ivec(psize(2)-ivsize+1:1:psize(2))=iv;

\end{document}